\documentclass[prl,twocolumn,english
,aps,reprint,superscriptaddress,notitlepage,nobibnote,preprintnumbers,showpacs]{revtex4-1}

\usepackage{graphicx}
\usepackage{dcolumn}
\usepackage{bm}

\usepackage{epsfig}
\usepackage{gensymb}
\usepackage{mathrsfs}
\usepackage{slashed}

\usepackage{amsmath}
\usepackage{amssymb}
\usepackage{mathtools}
\usepackage{graphicx,bm}
\usepackage{slashed}
\usepackage[makeroom]{cancel}
\usepackage{dsfont}
\usepackage{longtable}
\usepackage{multirow}
\usepackage{graphicx}
\usepackage{bbm}

\usepackage{lipsum, babel}
 \usepackage[titletoc]{appendix}

\def\fun#1#2{\lower3.6pt\vbox{\baselineskip0pt\lineskip.9pt
  \ialign{$\mathsurround=0pt#1\hfil##\hfil$\crcr#2\crcr\sim\crcr}}}

\newcommand\undermat[2]{
  \makebox[0.5pt][l]{$\smash{\underbrace{\phantom{%
    \begin{matrix}#2\end{matrix}}}_{ \let\scriptstyle\textstyle\text{\large $#1$}}}$}#2}
\newcommand\overmat[2]{
  \makebox[-1pt][l]{$\smash{\overbrace{\phantom{%
    \begin{matrix}#2\end{matrix}}}^{ \let\scriptstyle\textstyle\text{\large $#1$}}}$}#2}  
\newcommand*\circled[1]{ 
	\tikz[baseline= -2.5pt]{\node[shape=circle,draw,inner sep=0.3pt] (char) {#1};}}

\usepackage{tikz}
\usepackage{environ}
\makeatletter
\newsavebox{\measure@tikzpicture}
\NewEnviron{scaletikzpicturetowidth}[1]{%
  \def\tikz@width{#1}%
  \begin{lrbox}{\measure@tikzpicture}%
  \BODY
  \end{lrbox}%
  \pgfmathparse{#1/\wd\measure@tikzpicture}%
  \BODY
}
\makeatother
\usetikzlibrary{decorations.pathmorphing,decorations.markings,trees,shapes}
\usepackage[compat=1.1.0]{tikz-feynman}
\usepackage[vcentermath]{youngtab}
\usepackage{ytableau}

\def\lsim{\mathrel{\rlap{\raise 2.5pt \hbox{$<$}}\lower 2.5pt\hbox{$\sim$}}}
\def\gsim{\mathrel{\rlap{\raise 2.5pt \hbox{$>$}}\lower 2.5pt\hbox{$\sim$}}}

\newcommand{\bea}{\begin{eqnarray}}
\newcommand{\eea}{\end{eqnarray}}

\input epsf

\usepackage{color}

\newcommand{\comment}[1]{}

\newcommand{\mc}[1]{\mathcal{#1}}

\newcommand{\vev}[1]{\langle #1 \rangle}
\newcommand{\ket}[1]{| #1 \rangle}
\newcommand{\bra}[1]{\langle #1 |}

\newcommand{\eq}[1]{\begin{equation}\begin{split} #1 \end{split}\end{equation}}
\newcommand{\eqs}[1]{\begin{align} #1 \end{align}}

\renewcommand\arraystretch{2}

\begin{document}

\title{Depicting the Landscape of Generic Effective Field Theories}

\author{Hao-Lin Li}\email{lihaolin@itp.ac.cn}
\affiliation{
CAS Key Laboratory of Theoretical Physics, Institute of Theoretical Physics,
Chinese Academy of Sciences, Beijing 100190, China.}
\author{Jing Shu}\email{jshu@itp.ac.cn}
\affiliation{
CAS Key Laboratory of Theoretical Physics, Institute of Theoretical Physics,
Chinese Academy of Sciences, Beijing 100190, China.}
\affiliation{School of Physical Sciences, University of Chinese Academy of Sciences, Beijing 100049, P. R. China.}
\affiliation{CAS Center for Excellence in Particle Physics, Beijing 100049, China}
\affiliation{Center for High Energy Physics, Peking University, Beijing 100871, China}
\affiliation{School of Fundamental Physics and Mathematical Sciences, Hangzhou Institute for Advanced Study, UCAS, Hangzhou 310024, China}
\affiliation{International Centre for Theoretical Physics Asia-Pacific, Beijing/Hangzhou, China}
\author{Ming-Lei Xiao}\email{mingleix@itp.ac.cn}
\affiliation{
CAS Key Laboratory of Theoretical Physics, Institute of Theoretical Physics,
Chinese Academy of Sciences, Beijing 100190, China.}
\author{Jiang-Hao Yu}\email{jhyu@itp.ac.cn}
\affiliation{
CAS Key Laboratory of Theoretical Physics, Institute of Theoretical Physics,
Chinese Academy of Sciences, Beijing 100190, China.}
\affiliation{School of Physical Sciences, University of Chinese Academy of Sciences, Beijing 100049, P. R. China.}
\affiliation{School of Fundamental Physics and Mathematical Sciences, Hangzhou Institute for Advanced Study, UCAS, Hangzhou 310024, China}
\affiliation{International Centre for Theoretical Physics Asia-Pacific, Beijing/Hangzhou, China}

\begin{abstract}\indent
We describe a general framework to depict the space of Lorentz invariant effective field theories, which we call the landscape, given any kind of gauge symmetry and field content.
Various operator bases in the landscape can be systematically constructed, with the help of amplitude-operator correspondence, to emphasize different aspects: operator independence (y-basis), flavor relation (p-basis) and conserved quantum number (j-basis).
The transformation matrices among the bases encode model-independent properties of the operators, such as implications on their UV origin.
We illustrate this salient feature in examining the dimension 9 operators relevant for neutron-antineutron oscillation.
\end{abstract}

\maketitle

\section{Introduction}

Effective field theories (EFT) have been gradually gaining importance both theoretically and experimentally as a framework to describe the low energy effects of the ultraviolet (UV) physics in a model-independent way. 
The effective Lagrangians live in the space spanned by the operator basis $\mc{O}_i$, called the EFT landscape, while the Wilson coefficients $C_i$ of a certain theory is its coordinate in the landscape:
\begin{equation}\label{eq:Lag_EFT}
    \mc{L} = \sum_i C_i \mc{O}_i
    \in \bigoplus_{\rm type}\textrm{span}\left(\mc{O}_1, \cdots, \mc{O}_{\mathbbm{d}}\right).
\end{equation}
The word ``landscape" was already used in literature \cite{Elvang:2018dco,Elvang:2020lue} which, though, only examined quantum field theories with extra global symemtries, like conformal symmetry or low-energy theorems. In this letter, we are not assuming any of such symmetries, which leads to the whole landscape of ``generic EFTs'' as the direct sum of independent subspaces, called types, that consist of the same collection of fields and the same number of covariant derivatives. 
The EFTs with extra symmetries can be specified in a self-contained manner as some invariant subspaces of the renormalization group flows in the landscape. 

To study the landscape, the first subtlety is to find an operator basis that is complete and independent. There were several well-known redundancy relations among the operators: the equation of motion (EOM), integration-by-parts (IBP) and the flavor relations in the presence of repeated fields. These difficulties were either solved by brute force \cite{Grzadkowski:2010es,Liao:2016qyd,Murphy:2020rsh} as eliminating redundancy case by case, or circumvented by only counting the dimension $\mathbbm{d}$ for each type \cite{Henning:2015alf,Fonseca:2019yya}.
We surmount the subtleties by establishing the amplitude-operator correspondence~\cite{Ma:2019gtx,Shadmi:2018xan} and utilizing group theoretic techniques developed in ref.~\cite{Li:2020gnx,Li:2020xlh,Henning:2019enq, Henning:2019mcv} and provide a new and systematic method~\cite{Li:2020gnx,Li:2020xlh} to obtain
the dim-8 and dim-9 operator bases in the standard model effective field theory (SMEFT), as well as low energy effective field theory (LEFT)~\cite{Li:2020tsi} and  SMEFT with sterile neutrino ($\nu$SMEFT)~\cite{nuSMEFT}.

In this work, we formulate the framework of depicting the landscape, and apply it to generic EFTs with arbitrary gauge symmetries and field content. Besides the y-basis and p-basis defined in \cite{Li:2020gnx,Li:2020xlh}, we also extend the partial wave basis defined in ref.~\cite{Jiang:2020sdh} to a general operator j-basis taking into account the relevant gauge quantum numbers, which combined with the other two basis helps us gain insight on the UV origins of EFT operators.

\section{On-shell correspondence}

We start by defining the building blocks $D^w\Psi$ as $w\geq 0$ covariant derivatives acting on a field $\Psi$ of the Lorentz irreducible representation  $(j_l,j_r)$ and of the gauge group representation $\mathbf{r}$.
They are generally under reducible Lorentz representations decomposed as
\eq{\label{eq:building_block}
    D^w\Psi \in \Big(j_l+\frac{w}{2}, j_r+\frac{w}{2}\Big) \oplus \text{lower weights}
}
Assuming the particle state $\ket{\Psi}$ generated by the field is massless with helicity $h$, we can define the spinor helicity variables as usual $P_\mu\sigma^\mu_{\alpha\dot\alpha}\ket{\Psi} = \lambda_\alpha \tilde\lambda_{\dot\alpha}\ket{\Psi}$, so that the following matrix element of the building block
\eq{\label{eq:correspondence}
    \bra{0}D^w\Psi\ket{\Psi} \sim \lambda^{2j_l+w}\tilde\lambda^{2j_r+w}, 
}
with $(j_l,j_r) = (-h,0)$ for $h\le0$ and $(j_l,j_r) = (0,h)$ for $h\ge0$, transforms as the highest weight irreducible representation in eq.~\eqref{eq:building_block} due to the respective total symmetries among the dotted and undotted spinor indices.
It is proved in \cite{Li:2020gnx} that the lower weight representations of the building block can all be converted to other fields by using the EOM~\footnote{Regarding matrix element between physical states, the EOM is equivalent to zero.} or the $i[D,D]=F$ identity, which explains why they do not contribute in eq.~\eqref{eq:correspondence}. 
When massive particles are involved, we show in ref.~\cite{Li:2020tsi} that our method is still valid for massive scalars and fermions.

The dim-$d$ operators are constructed by combining the building blocks as
\eq{\label{eq:operator}
    & \mc{O}^{(d)} = T^{a_1,\dots,a_N}_G \epsilon^n \tilde\epsilon^{\tilde{n}} \prod_{i=1}^N D^w\Psi_{i,a_i}\ ,\\
    & \qquad \tilde{n}+n = d-N, \quad \tilde{n}-n = \sum_ih_i\ .
}
where $a_i$ are the indices for the representations $\mathbf{r}_i$ and $T$ is a gauge invariant tensor that contract all of them; $\epsilon^{\alpha\beta}$ and $\tilde\epsilon^{\dot\alpha\dot\beta}$ serve to contract all the spinor indices in $su(2)_l$ and $su(2)_r$ respectively.
Eq.~\eqref{eq:correspondence} induces the following form of matrix element
\eq{\label{eq:amp_basis}
    &\int d^4x\,\bra{0}\mc{O}^{(d)}(x)\ket{\Psi_1^{a_1},\dots,\Psi_N^{a_N}} \\
    &\quad \sim T^{a_1,\dots,a_N}_G\mc{B}^{(d)}(h_1,\dots,h_N)\delta^{(4)}(\textstyle{\sum_{i=1}^N}\lambda_i\tilde\lambda_i)
}
where $\mc{B} = \vev{\cdot}^n[\cdot]^{\tilde{n}}$ is constructed by applying the same $su(2)_{l,r}$ contractions as in eq.~\eqref{eq:operator} to the product of matrix elements in eq.~\eqref{eq:correspondence}, where the notations $\vev{ij}\equiv \lambda_i^\alpha\lambda_{j\alpha}$, $[ij]\equiv \tilde\lambda_{i\dot\alpha}\tilde\lambda_j^{\dot\alpha}$ are understood. 
The significance of the matrix element is that all the redundancy relations among operators lead to equal matrix elements:
1) operators differing by EOM has the same matrix element according to eq.~\eqref{eq:correspondence}; 
2) the integration over spacetime in eq.~\eqref{eq:amp_basis} further eliminates the IBP redundancy, which requires the total momentum of the matrix element to vanish as indicated by the $\delta$-function. 
The zero total momentum implies that by crossing symmetry, this matrix element could be interpreted as a local amplitude.
Therefore, we manage to formulate the amplitude-operator correspondence introduced in \cite{Ma:2019gtx} formally as eq.~\eqref{eq:amp_basis}, which maps the complete operator basis free from both EOM and IBP to an independent basis of local amplitudes, or an amplitude basis.

We define the set of Lorentz factors with the same helicities $(h_1,\dots,h_N)$ and at the same dimension $d$ as a \emph{Lorentz class}, and the set of group factors with the same constituting representations $(\mathbf{r}_1,\dots,\mathbf{r}_N)$ as a \emph{gauge class}. Given the classes, the space of possible Lorentz and gauge invariant structures in the local amplitudes is determined, which we call a \emph{type} of amplitudes. Since particles of the same helicities and gauge representations are involved equally in a type of amplitudes, we always put them in a flavor multiplet. Although amplitudes with external particles from the same flavor multiplet may be constraint by the spin-statistics theorem, it turns out to be inspiring to also study a type of amplitude structures without the constraint, which corresponds to a \emph{flavor-blind} operator where all the fields are distinguiable. In contrast, those amplitudes with the spin-statistics constraint correspond to the basis of \emph{flavor-specified} operators~\cite{Li:2020xlh,Liao:2019tep}.

\section{y-basis for flavor-blind operators}\label{sec:ybasis}

In this section, we construct an independent basis for a type of flavor blind operators, which is the direct product of the Lorentz basis $\{\mc{B}_i\}$ and the gauge basis $\{T_i\}$ for the given classes.

To find the independent set of Lorentz factors modulo the total momentum, ref.~\cite{Henning:2019enq, Henning:2019mcv} introduce an auxiliary $SU(N)$ transformation, which acts on $(\lambda_i,\tilde\lambda_i)$ such that the total momentum is invariant. 
As a result, the space of Lorentz factors with the same $(N,n,\tilde{n})$, where $n,\tilde{n}$ are defined in eq.~\eqref{eq:operator}, is decomposed into $SU(N)$ representation spaces in which either all or none of the Lorentz factors are proportional to the total momentum. After imposing momentum conservation, only the latter survive, which are specified as the following Young diagram,

\vspace{0.1cm}
\begin{eqnarray}\label{eq:primary_YD}
Y_{N,n,\tilde{n}} \quad = \quad \arraycolsep=0.15pt\def\arraystretch{1}
\rotatebox[]{90}{\text{$N-2$}} \left\{
    \begin{array}{cccccc}
        \yng(1,1) &\ \ldots{}&\ \yng(1,1)& \overmat{n}{\yng(1,1)&\ \ldots{}\  &\yng(1,1)} \\
        \vdotswithin{}& & \vdotswithin{}&&&\\
        \undermat{\tilde{n}}{\yng(1,1)\ &\ldots{}&\ \yng(1,1)} &&&
    \end{array}
\right.
 \\
\nonumber 
\end{eqnarray}
\vspace{0.1cm}

\noindent All the amplitudes of the form $\vev{\cdot}^n[\cdot]^{\tilde{n}}$ are represented by $SU(N)$ Young tableau generated by filling the labels $1,...,N$ into $Y_{N,n,\tilde{n}}$, where a 2-row column filled by $i,j$ represents a bracket $\vev{ij}$ and an $(N-2)$-row column filled by $k_1,\dots,k_{N-2}$ corresponds to $\mc{E}^{ijk_1,\dots,k_{N-2}}[ij]$ where $\mc{E}$ is the $SU(N)$ Levi-Civita.  
It was pointed out in \cite{Li:2020gnx} that the various Lorentz classes present in $Y_{N,n,\tilde{n}}$ correspond to Young tableau with different collections of filled-in labels $\{\underbrace{1,\dots,1}_{\#1},\underbrace{2,\dots,2}_{\#2},\dots \} $, where
\vspace{-3mm}
\eq{\label{eq:filling}
    \#i = \tilde{n}-2h_i .
}
For example, the Lorentz class $\psi_1\psi_2\psi_3\bar\psi_4 D$ has $(N,n,\tilde{n})=(4,2,1)$ according to eq.~\eqref{eq:operator}, while the labels to be filled in is determined by eq.~\eqref{eq:filling} as $\{1,1,2,2,3,3\}$. One immediately concludes that there is only one SSYT in this class:
\eq{\label{eq:lorentzexample}
    \young(112,233) \sim \mc{E}^{12ij}[ij]\vev{13}\vev{23} \sim \psi_1^{\alpha}\psi_2^{\beta}(D\psi_3)_{\alpha\beta\dot\alpha}\bar\psi_4^{\dot\alpha} \, .
}
where finally the Lorentz factor is converted back to the operator building blocks eq.~\eqref{eq:building_block} with the highest weight Lorentz representations.

In this letter, we assume that the gauge groups $G$ are all unitary groups.
With all the fields under tensor representations with only fundamental indices, the gauge factors $T_G$ are products of Levi-Civita tensors $\epsilon$'s represented by singlet Young tableau. An independent basis of them could be generated by applying the Littlewood-Richardson (L-R) rule invented in Ref.~\cite{Li:2020gnx}\footnote{The usual Littlewood-Richardson rule is for Young diagrams compared to ours for Young tableaux.} to the Young tableau representing each field.
For example, the tensor representation of a gluon field $G_{abc}$ in $\mathbf{8}$ of $SU(3)$ corresponds to $\scriptsize{\young(ab,c)}$, a quark field $q_d$ in $\mathbf{3}$ of $SU(3)$ corresponds to $\scriptsize{\young(d)}$, etc..
Hence the operator $G_{abc}q_dq_ed_f\bar{l}D$ with $SU(3)$ gauge class $(\mathbf{8}, \mathbf{3}, \mathbf{3}, \mathbf{3})$ have the following $SU(3)$ singlet Young tableau (detailed construction is provided in the appendix):
\eq{
& \young({{a}}{{b}},{{c}}{{d}},{{e}}{{f}}) \sim \epsilon^{ace}\epsilon^{bdf}=(T^{\rm y}_1)^{abcedf},\\
& \young({{a}}{{b}},{{c}}{{e}},{{d}}{{f}}) \sim \epsilon^{acd}\epsilon^{bef}=(T^{\rm y}_2)^{abcedf}.
}


Since both bases for the Lorentz and gauge factors are obtained as Young tableau, they are named \emph{y-basis}. Even without actual enumeration of them, their respective dimensions can be directly obtained by group theory techniques. 
For the Lorentz class, the dimension $\mathbbm{d}_{\mc{B}}$ is given by the following tensor product decomposition
\eq{\label{eq:lorentz_y_count}
    \bigotimes_{i=1}^N\ {\footnotesize \overmat{\mbox{\footnotesize{$\#i$}}}{\yng(2)\ \ldots{}\ \yng(1)} }  = \mathbbm{d}_{\mc{B}}\times \scalebox{0.6}{
    \arraycolsep=0.15pt\def\arraystretch{1}
\rotatebox[]{90}{\text{$N-2$}} $\left\{
    \begin{array}{cccccc}
        \yng(1,1) &\ \ldots{}&\ \yng(1,1)& \overmat{n}{\yng(1,1)&\ \ldots{}\  &\yng(1,1)} \\
        \vdotswithin{}& & \vdotswithin{}&&&\\
        \undermat{\tilde{n}}{\yng(1,1)\ &\ldots{}&\ \yng(1,1)} &&&
    \end{array}
\right.$
    } \oplus \dots\,,
}\vspace{1mm}
while for the gauge class the dimension $\mathbbm{d}_G$ is given by
\eq{\label{eq:gauge_y_count}
    \bigotimes_{i=1}^N\,\mathbf{r}_i = \mathbbm{d}_G\times\mathbf{1} + \dots
}
where $\mathbf{1}$ denotes the gauge singlet.
Taking the direct product of them, we obtain the y-basis $\mc{O}^{\rm y}=\mc{B}^{\rm y}\otimes T_G^{\rm y}$ for the flavor-blind operators of the given type. 
Any operator of the same type thus lives in the linear span of the rank-($\mathbbm{d} = \mathbbm{d}_{\mc{B}}\times\mathbbm{d}_G$) y-basis
\eq{\label{eq:coord}
    \mc{O} = \sum_{i=1}^{\mathbbm{d}}\mc{K}_i\mc{O}^{\rm y}_i \in \text{span}\left(\mc{O}^{\rm y}_1,\dots,\mc{O}^{\rm y}_{\mathbbm{d}}\right),
}
identified by a unique coordinate $(\mc{K}_1,\dots,\mc{K}_{\mathbbm{d}})$.
This coordinate can be found by the following steps \cite{Li:2020xlh}: 1) decompose each building block as in eq.~\eqref{eq:correspondence}, and convert all but the highest weight component into other types; 2) translate the Lorentz factors into Young tableau of the group $SU(N)$, and use the Fock conditions~\cite{book:Ma} to reduce them to the SSYT's, which amounts to using the Fierz identities and the IBP by a specific strategy elaborated in ref.~\cite{Li:2020xlh}; 3) with the numerical values of the gauge invariant tensors, define the inner product of gauge factors $(T_1, T_2) \equiv \sum_{a_1,a_2,...}T_1^{a_1a_2...}T_2^{a_1a_2...}$ that facilitates the projection of a tensor onto the y-basis gauge factors. 
The capability of finding the coordinate \eqref{eq:coord} is an essential feature of our framework, as will be clear soon.

\section{p-basis for flavor-specified operators}

After obtaining the complete basis for flavor-blind operators, we proceed to investigate the \emph{flavor-specified operators}. 
Naively, there should be $(\mathbbm{d} \times \prod_{i=1}^N n^{\rm f}_i)$ operators, where $n^{\rm f}_i$ is the number of flavors for the $i$th flavor multiplet.
Yet when some of the particles come from the same flavor multiplet, extra constraint reduces the number of independent operators.
The constraint can be understood from both sides of the correspondence \eqref{eq:amp_basis}:
\begin{itemize}
    \item In Ref.~\cite{Li:2020gnx}, the constraint is understood to be the flavor relations among operators with repeated fields. The flavor relations originate from the permutation symmetries of the flavor indices, which inherites from the symmetries of the Lorentz and gauge factors.
    
    \item In Ref.~\cite{Li:2020xlh}, the constraint is interpreted as the spin-statistics requirement for physical amplitudes involving identical particles.
    The total (anti-)symmetry of the physical amplitudes among the identical bosons (fermions) imposes constraint between the permutation symmetry of the coefficient flavor tensor, reducing the number $\prod_{i=1}^N n^{\rm f}_i$, and that of the flavor-blind basis, reducing the number $\mathbbm{d}$. 
\end{itemize}
Either way, we need a new amplitude basis with definite permutation symmetries, namely the \emph{p-basis}.

If we only need the number of amplitude basis satisfying spin-statistics, there is a quick way to count with the help of plethysm. Plethysm is a binary operation $\circled{p}$ of Young diagrams. When we apply it to an $SU(N)$ representation $\mathbf{r}$ and a $S_m$ representation $[\lambda]$, it gives the irreducible components in the decomposition of $\mathbf{r}^{\otimes m}$ that have symmetry $[\lambda]$ under the permutation of $S_m$, which is exactly what we want. Therefore eq.~(\ref{eq:lorentz_y_count},\ref{eq:gauge_y_count}) could be modified as \cite{Li:2020gnx}

\eqs{
    & \bigotimes_{i}\,{\footnotesize \overmat{\mbox{\footnotesize{$\#i$}}}{\yng(2)\ \ldots{}\ \yng(1)} }  \ \circled{p} \,[\lambda_i^{\mc{B}}]^{(\intercal)} = \mathfrak{n}_{[\lambda^{\mc{B}}]}\times \scalebox{0.6}{
    \arraycolsep=0.15pt\def\arraystretch{1}
\rotatebox[]{90}{\text{$N-2$}} $\left\{
    \begin{array}{cccccc}
        \yng(1,1) &\ \ldots{}&\ \yng(1,1)& \overmat{n}{\yng(1,1)&\ \ldots{}\  &\yng(1,1)} \\
        \vdotswithin{}& & \vdotswithin{}&&&\\
        \undermat{\tilde{n}}{\yng(1,1)\ &\ldots{}&\ \yng(1,1)} &&&
    \end{array}
\right.$
    } \oplus \dots, \label{eq:lorentz_p_count} \\
    & \bigotimes_{i}\,\mathbf{r}_i\ \circled{p} \,[\lambda_i^{G}] = \mathfrak{n}_{[\lambda^{G}]}\times \mathbf{1} \oplus \dots \label{eq:gauge_p_count}
}
where $i$ runs over particle species and $[\lambda_i] \vdash m_i$, a partition of integer $m_i$, is the symmetry under permutation of $m_i$ identical particles, with transpose ${}^\intercal$ needed in the presence of odd number $\tilde{n}$ of $\mc{E}$'s in $\mc{B}$ such as eq.~\eqref{eq:lorentzexample}. 
The multiplicities $\mathfrak{n}_{[\lambda^{\mc{B},G}]}$ of the target Young diagrams then give the number of representation spaces with permutation symmetry $[\lambda^{\mc{B},G}]=\{[\lambda_1^{\mc{B},G}],[\lambda_2^{\mc{B},G}],\dots\}$ for both factors.
The representation spaces of the local amplitudes are then obtained by the inner products of them, which induce the multiplicity $\mathfrak{n}_{[\lambda]}$ for $[\lambda] = \{[\lambda_1],[\lambda_2],\dots\}$, satisfying 
\eq{\label{eq:pdim_match}
    \mathbbm{d}=\sum_{[\lambda]}\mathfrak{n}_{[\lambda]}d_{[\lambda]} .
}
When the flavor indices are specified, the operators in a $[\lambda]$ representation space span the same $\bigotimes_i SU(n^{\rm f}_i)$ invariant subspace \cite{Fonseca:2019yya}, denoted by an irreducible flavor tensor $\mc{O}^{[\lambda]}_{f_1,\dots,f_N}$.
Hence we obtain the total number of flavor-specified operators of a type as 
\eq{\label{eq:Ntot}
    \mc{N}_{\rm tot} = \sum_{[\lambda]} \mathfrak{n}_{[\lambda]} \prod_i\mc{S}([\lambda_i]^{(\intercal)},n^{\rm f}_i)\,,
}
where $\mc{S}$ is the Hook content formula~\cite{book:Ma,Fonseca:2019yya} and tableau transpositions ${}^\intercal$ are imposed for fermions.
Note that $\mc{S}([\lambda],n^{\rm f}) = 0$ when $[\lambda]$ has more than $n_f$ rows, or $\lambda > n^{\rm f}$. In \cite{Li:2020gnx,Li:2020xlh} we call $\mc{O}^{[\lambda]}_{f_1,\dots,f_N}$ with non-zero components as a \emph{term}, the number of which is given by $\sum_{\lambda_i\le n_i^{\rm f}}\mathfrak{n}_{[\lambda]}$~\footnote{
The \emph{term} defined in our framework differ from the same terminology used in other literature such as \cite{Fonseca:2019yya,Murphy:2020rsh}. 
In those context, a term stands for a flavor tensor of operators with redundancies called flavor relations, which may consist of several irreducible representations, i.e. the direct sum of several terms in our sense. 
The number of terms is ambiguous as given in \cite{Fonseca:2019yya}, and even the upper bound is less than that in our interpretation.
}.
We reproduce the operator counting for the SMEFT up to dimension 15 with higher efficiency than \cite{Fonseca:2019yya,Henning:2015alf}. The following is the result for dimension 16 which was not obtained before.
\begin{table}[h]
\begin{tabular}{c||r|r|r}
    \hline
    $n^{\rm f}$     &   \#types     &   $\sum_{{\rm type}}$\#terms  &   $\sum_{\rm type}\mc{N}_{\rm tot}$ \\
    \hline
    1               &   137,342     &   83,106,786                  &   83,106,786                      \\
    2               &   140,480     &   200,319,358                 &   5,416,184,324                   \\
    3               &   140,514     &   223,269,714                 &   104,832,630,678                 \\
    $\infty$        &   140,514     &   225,957,161                 &   N/A \\
    \hline
\end{tabular}
\caption{Number of types, terms and flavor-specified operators in the dim-16 SMEFT. The $\infty$ in the last row means that number of terms saturates the upper bound if the $n^f$ is large enough to exceed the row numbers of any $[\lambda]$ obtained. 
}
\label{tab:counting}
\end{table}
%

The counting for the $[\lambda]$ representations in eq.~\eqref{eq:pdim_match} not only helps to obtian the numbers in table~\ref{tab:counting}, but also guides us to enumerate all the p-basis operators with the corresponding flavor permutation symmetries.
It is natural to use symmetrizers to obtain the explicit forms of the p-basis operators. The symmetrizers we select are those $b^{[\lambda]}_{x=1,\dots,d_{[\lambda]}}$ that span a minimal left ideal $\mc{L}^{[\lambda]}$ of the group algebra space $\tilde{S}_m$ forming an irreducible space of the $[\lambda]$ representation, such that for any operators $\mc{O}$, $\{b^{[\lambda]}_1\circ\mc{O},\dots,b^{[\lambda]}_{d_{[\lambda]}}\circ\mc{O}\}$, if not vanishing, also form a $[\lambda]$ representation space.

Then by traversing the symmetrizers through the y-basis $\mc{O}^{\rm y}_\zeta$, we find the coordinates of the non-zero results on the y-basis as in eq.~\eqref{eq:coord}, which facilitates selecting the independent ones $\zeta = i_1,\dots,i_{\mathfrak{n}_{[\lambda]}}$ therein to form a complete p-basis
\eq{\label{eq:Kpy}
    \mc{O}^{\rm p}_i \equiv \mc{O}^{[\lambda],\zeta}_x \equiv b^{[\lambda]}_x \circ \mc{O}^{\rm y}_\zeta = \sum_{j=1}^{\mathbbm{d}}\mc{K}^{\rm py}_{ij}\mc{O}^{\rm y}_j\ ,
}
where the indices $([\lambda],\zeta,x)$ are collectively represented by $i=1,\dots,\mathbbm{d}$ thanks to the relation eq.~\eqref{eq:pdim_match}. 
As explained above eq.~\eqref{eq:Ntot}, for flavor-specified operators, we can take $\mc{O}^{[\lambda],\zeta}_{f_1,\dots,f_N}\equiv(\mc{O}^{[\lambda],\zeta}_{x=1})_{f_1,\dots,f_N}$.
%

\section{j-basis as eigenbasis of conserved observables}

Another physically important way to organize the amplitude basis is to find the eigenbasis of conserved observables for subsets of the external particles. 
As pointed out in \cite{Jiang:2020sdh}, the eigenbasis of angular momentum for a 2-partition of external particles induces selection rules for loop integrals; 
in this letter, we extend the idea to more general partitions and investigate eigenbasis of both angular momentum and gauge quantum numbers, which makes the full classification of the UV origin of operators possible, and provides useful information to model builders.

\begin{figure}
    \centering
    \includegraphics[scale=0.7]{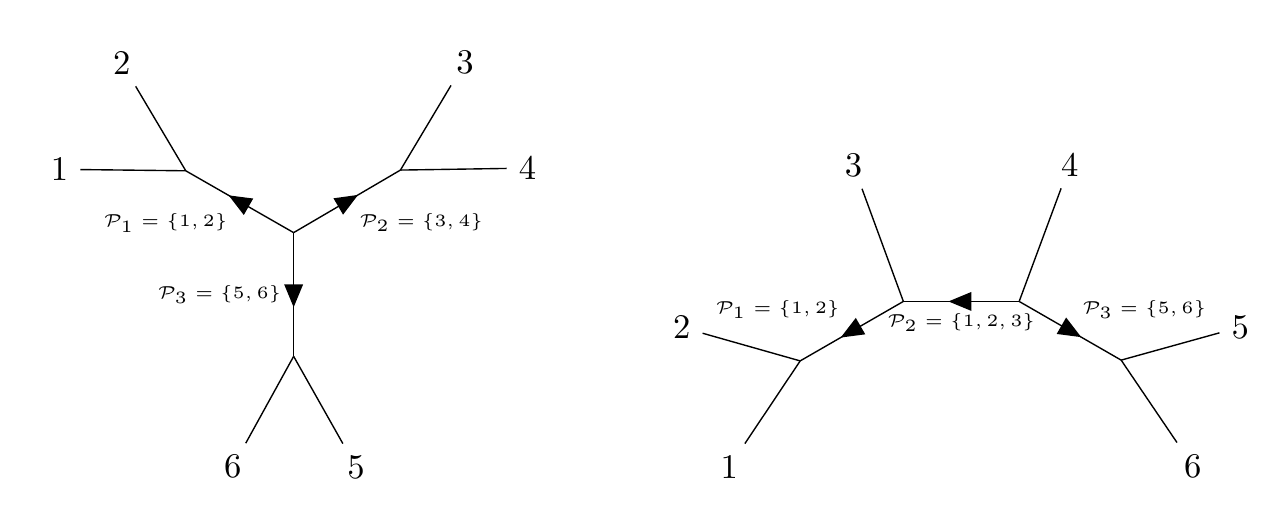}
    \caption{Two distinct 3-partitions of 6-point amplitudes specified by three subsets $\{\mc{P}_1,\mc{P}_2,\mc{P}_3\}$. The arrows suggest a clarifying way to associate each subset with an internal line that potentailly represents a resonance.}\label{fig:part}
\end{figure}

For an amplitude involving $N$ particles, at most $N-3$ non-trivial subsets $\mc{P}_k$ could be chosen to have commuting observables, satisfying: 1) each subset and its complement contains at least two particles; 2) for any two subsets, either one containes another or they do not intersect. Such a collecion of subsets is called a \emph{partition} $\mathbb{P}=(\mc{P}_1,\mc{P}_2, ..., \mc{P}_k)$, which defines a complete j-basis labelled by eigenvalues in each subset. 
In figure~\ref{fig:part} we show two possible topologies of 3-partitions for 6-point amplitudes.
For example, the SMEFT type $ d_c^\dagger{}^4u_c^\dagger{}^2$ has two inequivalent partitions of the first kind, namely $(d_c^\dagger d_c^\dagger,d_c^\dagger u_c^\dagger,d_c^\dagger u_c^\dagger)$ and $(d_c^\dagger d_c^\dagger,d_c^\dagger d_c^\dagger,u_c^\dagger u_c^\dagger)$, either of which determines a complete j-basis.

To define the angular momentum $J$ of a subset $\mc{P}$ in a covariant way, we need the Poincar\'e Casimir operator $W_\mc{P}^2$ in the spinor-helicity representation \cite{Jiang:2020sdh}, with eigenvalues $-s_{\mc{P}}J(J+1)$ where $s_{\mc{P}} = \left(\sum_{i\in\mc{P}}p_i\right)^2$.
It is straightforward to derive the eigenbasis of $W^2$, namely the Lorentz j-basis $\mc{B}^{\rm j}_{J,\dots}$, via diagonalizing its representation matrix. The ellipsis in the subscript denotes possible degeneracy.
By sequential diagonalizations of $W^2$ for all the subsets in a partition $\mathbbm{P}$, we could lift the degeneracy by the commuting angular momenta and obtain a subdivided eigenbasis,
\eq{
    & \mc{B}^{\rm j,\mathbbm{P}}_{(J)} \equiv \mc{B}^{\rm j}_{(J_1,\dots,J_k)_{\mathbbm{P}}} = (\mc{K}_{\mc{B}}^{\rm jy})_{(J_1,\dots,J_k)_{\mathbbm{P}}}^i \mc{B}^{\rm y}_i \ ,\\
    & s.t.\quad W_{\mc{P}_k}^2 \mc{B}^{\rm j,\mathbbm{P}}_{(J)} = -s_{\mc{P}_k} J_k(J_k+1)\mc{B}^{\rm j,\mathbbm{P}}_{(J)}\ ,\forall k .
}
together with the coordinates $\mc{K}_{\mc{B}}^{\rm jy}$ in the y-basis following eq.~\eqref{eq:coord}. 

The eigenbasis for the gauge groups are labeled by a tuple of irreducible representations for each subset $\mc{P}$: $(\mathbf{r})\equiv(\mathbf{r}_1, \mathbf{r}_2,.., \mathbf{r}_k)_{\mathbbm{P}}$.
Each of $\mathbf{r}_k$ should be accessible by the constituting particles in the subset $\mc{P}_k$, while the total direct product $\bigotimes_k\mathbf{r}_k$ should include a singlet representation due to the gauge symmetry.
For a given tuple $(\mathbf{r})$, one can impose a series of Young symmetrizers on the y-basis, which either vanishes or induces a j-basis gauge factor
\eq{
  T^{\rm j,\mathbbm{P}}_{(\mathbf{r}),i}\equiv T^{(\mathbf{r}_1,...,\mathbf{r}_k)}_{(x_1,...,x_k,\xi)} =\mc{Y}_{x_1}^{\mathbf{r}_1}\circ...\circ \mc{Y}_{x_k}^{\mathbf{r}_k}\circ T^y_\xi,
}
where $i$ represents a collection of indices $(x_1,...x_k,\xi)$ and $x_i$ denotes a specific Young tableaux representing $\mathbf{r}_i$ constructed via the L-R rule mentioned in the previous section.
Independent ones could be selected from the results by using the inner product of the tensors mentioned previously, and the total number of independnet group factors $\mathfrak{n}_{(\mathbf{r})}$ should add up to the dimension of the gauge y-basis: $\mathbbm{d}_G=\sum_{(\mathbf{r})} \mathfrak{n}_{(\mathbf{r})}$.  
The complete j-basis hence consists of independent combinations of the y-basis, with conversion matrix $\mc{K}^{\rm jy}_G$ given by
\eq{
T^{\rm j,\mathbb{P}}_{(\mathbf{r},i)}=(\mc{K}^{\rm jy}_G)^l_{(\mathbf{r},i)} T^{\rm y}_l.
}
A concrete example is illustrated in the appendix.

The full j-basis operaotrs $\mc{O}^{\rm j,\mathbbm{P}}_{(\mathbf{r},J)} $ corresponding to amplitudes $\sim T^{\rm j,\mathbbm{P}}_{(\mathbf{r})}\mc{B}^{\rm j,\mathbbm{P}}_{(J)}$ is another choice of complete operator basis, with linear transformation from the y-basis given by $\mc{K}^{\rm jy}=\mc{K}^{\rm jy}_G\otimes\mc{K}^{\rm jy}_{\mc{B}}$. It is important to express the physical amplitudes in terms of p-basis as combinations of the j-basis
\eq{\label{eq:convert_pj}
    \mc{O}^{\rm p} = \mc{K}^{\rm pj}\mc{O}^{\rm j} \equiv \mc{K}^{\rm py}(\mc{K}^{\rm jy})^{-1}\mc{O}^{\rm j},
}
which encodes the exact combinations of Wilson coefficients $C_{(\mathbf{r},J)} = \sum_\zeta C_\zeta \mc{K}^{\rm pj}_{\zeta,(\mathbf{r},J)}$
that contribute to processes with the given quantum numbers $(\mathbf{r},J)$. In \cite{Jiang:2020sdh}, such combinations show up as expected in the anomalous dimension matrix as a result of selection rules. 
We could also examine UV models by studying the j-basis, because they are the directly produced operators in a process with quantum numbers fixed by certain heavy resonances. To get their projections on the physical p-basis, we derive the inverse of $\mc{K}^{\rm pj}$\footnote{When taking the inverse of $\mc{K}^{\rm pj}$, we are using the full $\mathbbm{d}\times\mathbbm{d}$ matrix, even if some of them are not relevant in the complete basis of flavor-specified operators.},
\eq{\label{eq:inverse_pj}
    \mc{O}^{\rm j,\mathbbm{P}}_{(\mathbf{r},J)} = \sum_\zeta (\mc{K}^{\rm pj})^{-1}_{(\mathbf{r},J),\zeta} \mc{O}^{\rm p}_\zeta
}
which provides the combinations of p-basis operators that are generated by a particular UV resonance. 
These will be demonstrated in details by an explicit example next section.

\section{Example: $n$-$\bar{n}$ oscillation}

We take the neutron-anti-neutron $n-\bar{n}$ oscillation as an example for all the workflow we described.
This exotic process has baryon number violation $\Delta B=2$ and starts at dimension 9 in the SMEFT, whose leading contributions come from 3 types of operators
\eq{
    & d^\dagger_c{ }^4 u^\dagger_c{ }^2 \sim T^{abcdef}_{SU(3)}\psi^\dagger{ }^6,\\
    & Q^4d_c^{\dagger 2} \sim T^{abcdef}_{SU(3)}T^{ijkl}_{SU(2)}\psi^4\psi^{\dagger2}, \\
    & Q^{2} d^\dagger_c{ }^3 u^\dagger_c{ }^2  \sim T^{abcdef}_{SU(3)}T^{ij}_{SU(2)}\psi^2\psi^{\dagger4}.
}
The Lorentz classes involved are
\begin{equation}
\begin{array}{c|cc}
    \hline
    \text{Lorentz}  &  \multicolumn{2}{c}{\text{y-basis}} \\
    \hline
    \multirow{3}{*}{$\psi^6$}  
        & \mc{B}_1= \vev{12}\vev{34}\vev{56} 
        & \mc{B}_2= \vev{12}\vev{35}\vev{46} \\
        & \mc{B}_3= \vev{13}\vev{24}\vev{56}
        & \mc{B}_4= \vev{13}\vev{25}\vev{46} \\
        & \mc{B}_5= \vev{14}\vev{25}\vev{36}&\\
    \hline
    \psi^4\psi^{\dagger2}
        & \mc{B}_1= \vev{12}\vev{34}[56]
        & \mc{B}_2= \vev{13}\vev{24}[56] \\
    \hline
\end{array}
\end{equation}
The gauge classes involved are
\begin{equation}
\begin{array}{c|cc}
    \hline
    \text{gauge classes}  &  \multicolumn{2}{c}{\text{y-basis}} \\
    \hline
    \multirow{3}{*}{$T_{SU(3)}^{abcdef}$}  
        & T_{1}= \epsilon^{ace}\epsilon^{bdf}
        & T_{2}= \epsilon^{acd}\epsilon^{bef} \\
        & T_{3}= \epsilon^{abe}\epsilon^{cdf}
        & T_{4}= \epsilon^{abd}\epsilon^{cef} \\
        & T_{5}= \epsilon^{abc}\epsilon^{def}&\\
    \hline
    T_{SU(2)}^{ijkl}
        & T'_{1}= \epsilon^{ij}\epsilon^{kl}
        & T'_{2}= \epsilon^{ik}\epsilon^{jl} \\
    \hline
    T_{SU(2)}^{ij}
        & T'_{1}= \epsilon^{ij}& \\
    \hline
\end{array}
\end{equation}
Hence there are $\mathbbm{d}=25,20,10$ independent flavor-blind operators for the three types.

While we have derived the full p-basis summarized in table~\ref{tab:p-basis}, 
\begin{table}
\eq{\begin{array}{c|l}
    \text{type} & \quad\quad\bigoplus_{[\lambda]}\mathfrak{n}_{[\lambda]}\{[\lambda_1],[\lambda_2],\dots\} \\
    \hline
    \multirow{3}{*}{$d_c^{\dagger 4}u_c^{\dagger 2}$}
    & {\color{red} 2\{\scriptsize{\yng(2)}_u,\scriptsize{\yng(4)}_d\}} \oplus\, \{\scriptsize{\yng(2)}_u,\scriptsize{\yng(3,1)}_d\} \oplus\\
    & 2\,\{\scriptsize{\yng(2)}_u,\scriptsize{\yng(2,2)}_d\} \oplus\,
    2\{\scriptsize{\yng(1,1)}_u,\scriptsize{\yng(3,1)}_d\} \oplus\\
    & \{\scriptsize{\yng(2)}_u,\scriptsize{\yng(2,1,1)}_d\} \oplus\,
    2\{\scriptsize{\yng(1,1)}_u,\scriptsize{\yng(2,1,1)}_d\} \oplus\,
    \{\scriptsize{\yng(2)}_u,\scriptsize{\yng(1,1,1,1)}_d\} \\
    \hline
    \multirow{2}{*}{$Q^4d_c^{\dagger2}$}
    & {\color{red}\{\scriptsize{\yng(4)}_Q,\scriptsize{\yng(2)}_{d^\dagger}\}} \oplus\,
    3\,\{\scriptsize{\yng(2,2)}_Q,\scriptsize{\yng(2)}_{d^\dagger}\} \oplus\\
    & 2\,\{\scriptsize{\yng(3,1)}_Q,\scriptsize{\yng(1,1)}_{d^\dagger}\} \oplus\,
    2\,\{\scriptsize{\yng(2,1,1)}_Q,\scriptsize{\yng(1,1)}_{d^\dagger}\} \oplus\,
    \{\scriptsize{\yng(1,1,1,1)}_Q,\scriptsize{\yng(2)}_{d^\dagger}\} \\
    \hline
    \multirow{3}{*}{$Q^2d_c^{\dagger 3}u_c^\dagger$}
    & {\color{red}\{\scriptsize{\yng(2)}_{Q},\scriptsize{\yng(3)}_{d^\dagger}\}} \oplus\,
    2\,\{\scriptsize{\yng(2)}_{Q},\scriptsize{\yng(2,1)}_{d^\dagger}\} \oplus\\
    & \{\scriptsize{\yng(1,1)}_{Q},\scriptsize{\yng(3)}_{d^\dagger}\} \oplus\,
    \{\scriptsize{\yng(1,1)}_{Q},\scriptsize{\yng(2,1)}_{d^\dagger}\} \oplus\\
    & \{\scriptsize{\yng(2)}_{Q},\scriptsize{\yng(1,1,1)}_{d^\dagger}\} \oplus\,
    \{\scriptsize{\yng(1,1)}_{Q},\scriptsize{\yng(1,1,1)}_{d^\dagger}\} \\
    \hline
\end{array}\notag}
\caption{The allowed permutation symmetries among the flavors of the repeated fields and multiplicities of the corresponding representations in the full p-basis for 3 types of dim-9 operators relevant for $n$-$\bar{n}$ ocsillation in the SMEFT. 
The single fermion generation selects the totally symmetric representions colored in red.
}\label{tab:p-basis}
\end{table}
we are mostly interested in the operators involving quarks of first generation which have the leading contributions to the $n$-$\bar{n}$ oscillation. They reside only in the operators with totally symmetric flavors among the repeated fields, colored red in table~\ref{tab:p-basis}, which are explicitly given as
\eq{\label{eq:single_gen}
    &\mc{O}_1 = \epsilon^{ace}\epsilon^{bdf}(d^\dagger_{_c a}d^\dagger_{_c b})(d^\dagger_{_c c}d^\dagger_{_c d})(u^\dagger_{_c e}u^\dagger_{_c f})\ ,\\
    &\mc{O}_2 = \epsilon^{acd}\epsilon^{bef}(d^\dagger_{_c a}d^\dagger_{_c b})(d^\dagger_{_c c}u^\dagger_{_c e})(d^\dagger_{_c d}u^\dagger_{_c f})\ ,\\
    &\mc{O}_3 = \epsilon^{ace}\epsilon^{bdf}\epsilon^{ik}\epsilon^{jm}(Q_{ai}Q_{ck})(Q_{bj}Q_{dm})(d^\dagger_{_c e}d^\dagger_{_c f})\ ,\\
    &\mc{O}_4 = \epsilon^{aef}\epsilon^{bcd}\epsilon^{ij}(d^\dagger_{_c a}d^\dagger_{_c c})(d^\dagger_{_c b}u^\dagger_{_c d})(Q_{ei}Q_{fj})\ .
}

For the type $d_c^{\dagger 4}u_c^{\dagger 2}$, with operators identified as $\mc{O}=\mc{K}_\zeta^{ij}T_{SU(3),i}\mc{B}_j$, the p-basis with total flavor symmetry is given by the following coordinate matrices,
\eq{\label{eq:p_coord}\arraycolsep=1pt\def\arraystretch{1}
    \mc{K}_1 = \left(\begin{array}{ccccc}
    0 & -1 & 0 & 0 & 1 \\
    0 & -2 & 0 & 0 & 0 \\
    0 & 0 & 0 & 0 & -2 \\
    0 & 0 & 0 & -1 & 1 \\
    0 & 0 & 0 & -2 & 0
    \end{array}\right)\ 
    \mc{K}_2 = \left(\begin{array}{ccccc}
    -2 & 1 & 0 & 0 & 1 \\
    0 & 0 & 0 & 0 & 0 \\
    0 & 0 & 0 & 0 & 0 \\
    0 & 0 & -2 & 1 & -1 \\
    0 & 0 & 0 & 0 & 0
    \end{array}\right) .
}
which induce the operators $\mc{O}_{1,2}$ in eq.~\eqref{eq:single_gen} after taking all flavors to be the first generation. 
As such, we reproduce the result in ref.~\cite{Nussinov:2001rb} easily, without prolonged discussions about the redundancy relations. Moreover, 
the p-basis operators with other flavor symmetries $[\lambda]$ that involve at least two generations are simultaneously obtained, which are presented in \cite{Li:2020xlh}.
\begin{figure}
    \centering
    \includegraphics[scale=0.8]{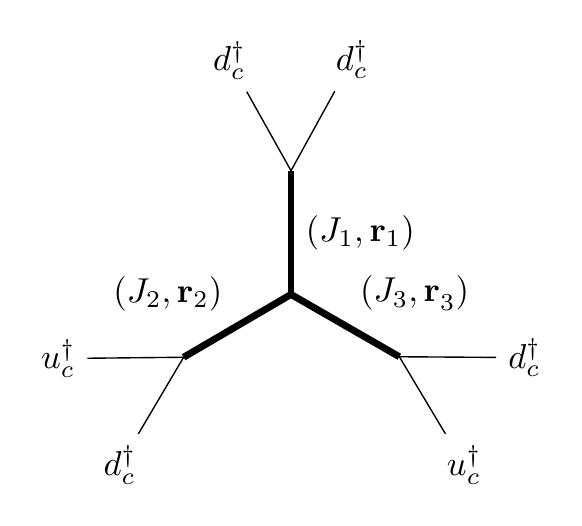}
    \caption{The possible UV digram generating  $d_c^{\dagger 4}u_c^{\dagger 2}$ operators. It involves three possible resonances with quantum numbers $(\mathbf{r}_i,J_i)$, $i=1,2,3$ that couple with $\mc{P}_1=(d_c^\dagger d_c^\dagger)$ and $\mc{P}_{2,3}=(d_c^\dagger u_c^\dagger)$.}\label{fig:topo}
\end{figure}

As an example, assume that $\mc{O}_{1,2}$ are generated by the process in figure~\ref{fig:topo} after integrating out the heavy particles denoted by the thick lines. To infer the quantum number of those heavy fields, 
we investigate the j-basis for the partition $\mathbbm{P}=(\mc{P}_1,\mc{P}_2,\mc{P}_3)=(d_c^\dagger d_c^\dagger,d_c^\dagger u_c^\dagger,d_c^\dagger u_c^\dagger)$.
We have the conversion matrix for the $SU(3)$ factor $\mc{K}^{\rm jy}_{SU(3)}$:
\begin{eqnarray}\arraycolsep=1pt\def\arraystretch{1}
   \left( \begin{array}{c}
  T^{(\mathbf{6},\mathbf{6},\mathbf{6})}\\ 
  T^{ (\mathbf{6},\bar{\mathbf{3}},\bar{\mathbf{3}}) }\\
  T^{ (\bar{\mathbf{3}}, \mathbf{6},\bar{\mathbf{3}}) } \\
   T^{  (\bar{\mathbf{3}}, \bar{\mathbf{3}},{\mathbf{6}})  }\\
   T^{  (\bar{\mathbf{3}}, \bar{\mathbf{3}},\bar{\mathbf{3}})  }
 \end{array}\right)
 =\left(\begin{array}{ccccc}
      -1/2 & 1 & 1/4 & -1/2 & 3/4\\
      0 & 1/2 & 0 & -1/4 &1/4 \\
      0 & 0 & -1/4 & 0 & 1/4\\
      0 & 0 & -1/4 & 1/2 & -1/4 \\
      0 & 0 & 1/2 & 0 & 1/2
 \end{array}\right) 
    \left( \begin{array}{c}
  T_1\\ 
  T_2\\
  T_3 \\
   T_4\\
   T_5
 \end{array}\right)\nonumber \\
\end{eqnarray}
and that for the Lorentz factor $\mc{K}^{\rm jy}_{\mc{B}}$
\eq{\arraycolsep=1pt\def\arraystretch{1}
\left(\begin{array}{c}
\mc{B}^{(1,1,1)} \\ \mc{B}^{(0,1,1)} \\ \mc{B}^{(1,0,1)} \\ \mc{B}^{(1,1,0)} \\ \mc{B}^{(0,0,0)}
\end{array}\right) = \left(\begin{array}{ccccc}
0 & 0 & 1 & -1 & 1 \\
-2 & 1 & 0 & 0 & 0 \\
2 & -1 & -2 & 0 & 2 \\
0 & 1 & 0 & -2 & 0 \\
0 & 1 & 0 & 0 & 0
\end{array}\right) \left(\begin{array}{c}
\mc{B}_1 \\ \mc{B}_2 \\ \mc{B}_3 \\ \mc{B}_4 \\ \mc{B}_5
\end{array}\right)
}
We could derive the coordinates of the two operators eq.~\eqref{eq:p_coord} in the j-basis following eq.~\eqref{eq:convert_pj},
\eq{\label{eq:Kpj}
    \mc{K}^{\rm pj}_{\zeta,(\mathbf{r},J)} = \mc{K}_\zeta^{ij}(\mc{K}_{SU(3)}^{\rm jy})^{-1}_{i,(\mathbf{r})}(\mc{K}_{\mc{B}}^{\rm jy})^{-1}_{j,(J)} ,
}
as their contributions to processes with various quantum numbers.
On the other hand, to study the UV origins of the two operators, we take the inverse of eq.~\eqref{eq:Kpj} to obtain $(\mc{K}^{\rm p j})^{-1}$ in eq.~\eqref{eq:inverse_pj}
and find the combinations of operators generated by the diagram in figure~\ref{fig:topo} with various quantum numbers.
\begin{table}[h]\scriptsize
    \eq{\begin{array}{c|ccccc}
     (\mathbf{r}_i,J_i)                                     &   (1,1,1)             &   (0,1,1)             &   (1,0,1)             &   (1,1,0)             &   (0,0,0) \\
    \hline
    (\mathbf{6},\mathbf{6},\mathbf{6})                  &   0                   &  3\mc{O}_1+8\mc{O}_2  &   0                   &   0                   &   \mc{O}_1-8\mc{O}_2       \\
    (\mathbf{6},\bar{\mathbf{3}},\bar{\mathbf{3}})      &   0                   &  \mc{O}_2             &   0                   &   0                   &   \mc{O}_2   \\
    (\bar{\mathbf{3}},\mathbf{6},\bar{\mathbf{3}})      &   3\mc{O}_1+8\mc{O}_2 &   0                   &  \mc{O}_1-8\mc{O}_2   &  3\mc{O}_1+8\mc{O}_2  &   0         \\
    (\bar{\mathbf{3}},\bar{\mathbf{3}},\mathbf{6})      &   3\mc{O}_1+8\mc{O}_2 &   0                   &  3\mc{O}_1+8\mc{O}_2  &  \mc{O}_1-8\mc{O}_2   &   0         \\
    (\bar{\mathbf{3}},\bar{\mathbf{3}},\bar{\mathbf{3}})&  -3\mc{O}_1+8\mc{O}_2 &   0                   &  3\mc{O}_1+8\mc{O}_2  &  3\mc{O}_1+8\mc{O}_2  &   0   
    \end{array}\notag}
    \caption{Each entry is associated to the three heavy resonances in fig.~\ref{fig:topo} with a particular combination of $(\mathbf{r}_i,J_i)$. Specific combinations of the operators $\mc{O}_{1,2}$ that could be generated by the resonances are given up to constant factors. }\label{tab:O_comb}
\end{table}
The zero entries, implying no operators are generated for $(\mathbf{r}_1,J_1)=(\mathbf{6},1)$ or $(\bar{\mathbf{3}},0)$, results from the permutation symmetry among the subset $\mc{P}_1=(d_c^\dagger d_c^\dagger)$ that violates the spin-statatistics of the two identical fermions, according to the selection rule B in \cite{Jiang:2020sdh}. 
Another constraint comes from the identity between the two parts $\mc{P}_2=\mc{P}_3=(d_c^\dagger u_c^\dagger)$, which means the permutation of them should be a symmetry. Indeed, the permutation swaps $(\mathbf{r}_2,J_2)$ and $(\mathbf{r}_3,J_3)$, which amounts to swapping the 3rd and 4th rows and also the 3rd and 4th columns in table~\ref{tab:O_comb}, reflecting the symmetry therein.
In addition, table~\ref{tab:O_comb} further shed the light on the UV origin of $\mc{O}_{1,2}$: 1) If the heavy fields in the UV model coupling to $d_c^\dagger d^\dagger_c$ and $u_c^\dagger d_c^\dagger$ are in $\mathbf{6}$ and $\bar{\mathbf{3}}$ of $SU(3)$ respectviely, then the model cannot generate $\mc{O}_1$ as can be observed in the second line in table~\ref{tab:O_comb}; 2) If $J_1=J_2=J_3=0$, the allowed scenario with three color sextets $SU(2)$ singlets scalars is studied in Ref.~\cite{Babu:2006xc}, while the scenario $\mathbf{r}_1=\mathbf{6},\mathbf{r}_2=\mathbf{r}_3=\bar{\mathbf{3}}$ is still unexplored; 3) If $J_1=J_2=J_3=1$, then two  $\bar{\mathbf{3}}$  fields must exist that couple to $d_c^\dagger d_c^\dagger$  and   $d_c^\dagger u_c^\dagger$ respectively; 
4) In the scenario that two $u_c^\dagger d_c^\dagger$ pairs couple the same heavy fields, the ratio of the Wilson coefficients of $\mc{O}_1$ and $\mc{O}_2$ can be predicted as in table~\ref{tab:O_comb}.
Finally, we comment that the vector di-quark model in Ref.~\cite{Assad:2017iib} can also contribute to this topology but with one of the $u_c^\dagger$ and one of $d_c^\dagger$ changed to left handed doublet $Q$'s.

\section{Conclusion and Outlook}

We proposed a general framework to construct the independent and complete operator bases for Lorentz invariant EFTs with  arbitrary gauge symmetries and field content at any mass dimension. In our prescription, no flavor relations are necessary for operators with repeated fields.
Although we only demonstrate the power of this framework in the SMEFT up to dimension 16, it could be easily applied to other EFTs including, but not limited to, LEFT, $\nu$SMEFT, two Higgs doublet model, left-right symmetry, composite Higgs, and dark matter EFT, etc. 

The various operator bases of the landscape we obtained bring us a lot of benefits. 
Firstly, one could study all kinds of exotic processes, whose leading contributions appear at high dimensions, regarding their UV origins in a model-independent way. Besides the $n$-$\bar{n}$ oscillation discussed in this work, there are more processes such as the proton exotic decay, neutrinoless double beta decay, and electric dipole moment, etc. to be explored. As demenstrated from the $n$-$\bar{n}$ oscillation, based on the j-basis, it is possible to list various possible UV contributions for a specific processes.
Another application is to compute the renormalization group flow in an efficient way, thanks to the selection rules for the anomalous dimension matrix \cite{Jiang:2020sdh,Baratella:2020dvw}. 
Furthermore, features of the renormalization group flow, such as attractors and fixed points, could be identified in the landscape as EFTs with extra global symmetries.

\section{Acknowledgements}
H.L.L., M.L.X. and J.H.Y. thank Feng-Kun Guo and Bing-Song Zou for helpful discussion. H.L.L. and J.H.Y. were supported by the National Science Foundation of China (NSFC) under Grants No. 11875003 and No. 12022514. J.S. is supported by the National Natural Science Foundation of China under Grants No. 11947302, No. 11690022, No. 11851302, No. 11675243, and No. 11761141011; and is supported by the Strategic Priority Research Program and Key Research Program of Frontier Science of the Chinese Academy of Sciences under Grants No. XDB21010200, No. XDB23000000, and No. ZDBS-LY-7003.  
M.L.X. is supported by the National Natural Science Foundation of China (NSFC) under grant No.2019M650856 and the 2019 International Postdoctoral Exchange Fellowship Program.
JHY was also supported by the National Science Foundation of China (NSFC) under Grants No. 11947302.

\appendix
\section{An Example of Gauge j-basis}
Let us derive the Gauge j-basis for the operator type $G q^2 d \bar{l}D$, where $G$ is the field strength tensors for the $SU(3)$ gauge group, $q$ is the left-handed quark doublet, $d$ is the right-handed down quark, $l$ is the left-handed letpon double, $D$ represents derivative.
Let us first lable the $SU(3)$ indices and Young tableaux of the fields in the following:
\begin{eqnarray}
	&\epsilon_{acd}\lambda^A{}_b^d G^A = G_{abc} \sim \young(ab,c),\\
	& \quad q_d\sim \young(d)\quad q_e\sim \young(e)\quad d_f\sim \young(f),
\end{eqnarray}
where we convert the adjoint represeentation index of $G$ to the fundamental ones. The y-basis is obtained by the modified L-R rule:
\begin{eqnarray}
&& \young({{a}}{{b}},{{c}})\xrightarrow{\young({{d}})}\young({{a}}{{b}},{{c}}{{d}})\xrightarrow{\young({{e}})}\young({{a}}{{b}},{{c}}{{d}},{{e}})\xrightarrow{\young({{f}})}\young({{a}}{{b}},{{c}}{{d}},{{e}}{{f}}) \sim \epsilon^{ace}\epsilon^{bdf}=T^{\rm y}_1,\nonumber \\
&& \young({{a}}{{b}},{{c}})\xrightarrow{\young({{d}})}\young({{a}}{{b}},{{c}},{{d}})\xrightarrow{\young({{e}})}\young({{a}}{{b}},{{c}}{{e}},{{d}})\xrightarrow{\young({{f}})}\young({{a}}{{b}},{{c}}{{e}},{{d}}{{f}}) \sim \epsilon^{acd}\epsilon^{bef}=T^{\rm y}_2.\nonumber \\
\end{eqnarray}  
For the partition $\mc{P} = (Gq)$, the accesible irreducible representation are $\bar{\mathbf{6}}$ and $\mathbf{3}$, with corresponding Young tableaux:
\begin{eqnarray}
\young({{a}}{{b}},{{c}}{{d}})\ ,\quad \young({{a}}{{b}},{{c}},{{d}})\ .
\end{eqnarray}

Then one can act the Young symmetrizers for these tableaux to the y-basis to obtain the corresponding j-basis 
\eq{
T^{\bar{\mathbf{6}}} & \equiv\mathcal{Y}\left[\scriptsize{\young({{a}}{{b}},{{c}}{{d}})}\right] \circ (T^y_1)^{\bm{a}\bm{b}\bm{c}\bm{d}ef} \\
&= \frac{1}{3}\left(\epsilon^{ace}\epsilon^{bdf}+\epsilon^{acf}\epsilon^{bde}\right) \\
& \qquad\quad + \frac{1}{6}\left(\epsilon^{adf}\epsilon^{bce}+\epsilon^{ade}\epsilon^{bcf}+\epsilon^{abf}\epsilon^{cde}+\epsilon^{abe}\epsilon^{cdf}\right) \\
&= \frac{3}{4}T^{\rm y}_1-\frac{1}{4}T^{\rm y}_2,\\
T^{\mathbf{3}} & \equiv\mathcal{Y}\left[\scriptsize{\young({{a}}{{b}},{{c}},{{d}})}\right] \circ (T^y_1)^{\bm{a}\bm{b}\bm{c}\bm{d}ef} \\
&= \frac{1}{4}\left(\epsilon^{adf}\epsilon^{bce}-\epsilon^{ade}\epsilon^{bcf}-\epsilon^{acf}\epsilon^{bde}+\epsilon^{acd}\epsilon^{bdf}\right. \\
& \qquad\quad - \left.\epsilon^{abf}\epsilon^{cde}+\epsilon^{abe}\epsilon^{cdf}\right) \\
&=\frac{1}{2}T^{\rm y}_2.
}
Since we know a priori that the multiplicities of these two irreps are 1, we have already obtained the complete set of j-basis, which can also be verified by
\begin{eqnarray}
\mathcal{Y}\left[\scriptsize{\young({{a}}{{b}},{{c}}{{d}})}\right] \circ T^{\rm y}_2=0,\quad \mathcal{Y}\left[\scriptsize{\young({{a}}{{b}},{{c}},{{d}})}\right] \circ T^{\rm y}_2 = 2T^{\mathbf{3}}.
\end{eqnarray}

\bibliographystyle{apsrev4-1}
\bibliography{ref}

\begin{thebibliography}{26}%
\makeatletter
\providecommand \@ifxundefined [1]{%
 \@ifx{#1\undefined}
}%
\providecommand \@ifnum [1]{%
 \ifnum #1\expandafter \@firstoftwo
 \else \expandafter \@secondoftwo
 \fi
}%
\providecommand \@ifx [1]{%
 \ifx #1\expandafter \@firstoftwo
 \else \expandafter \@secondoftwo
 \fi
}%
\providecommand \natexlab [1]{#1}%
\providecommand \enquote  [1]{``#1''}%
\providecommand \bibnamefont  [1]{#1}%
\providecommand \bibfnamefont [1]{#1}%
\providecommand \citenamefont [1]{#1}%
\providecommand \href@noop [0]{\@secondoftwo}%
\providecommand \href [0]{\begingroup \@sanitize@url \@href}%
\providecommand \@href[1]{\@@startlink{#1}\@@href}%
\providecommand \@@href[1]{\endgroup#1\@@endlink}%
\providecommand \@sanitize@url [0]{\catcode `\\12\catcode `\$12\catcode
  `\&12\catcode `\#12\catcode `\^12\catcode `\_12\catcode `\%12\relax}%
\providecommand \@@startlink[1]{}%
\providecommand \@@endlink[0]{}%
\providecommand \url  [0]{\begingroup\@sanitize@url \@url }%
\providecommand \@url [1]{\endgroup\@href {#1}{\urlprefix }}%
\providecommand \urlprefix  [0]{URL }%
\providecommand \Eprint [0]{\href }%
\providecommand \doibase [0]{http://dx.doi.org/}%
\providecommand \selectlanguage [0]{\@gobble}%
\providecommand \bibinfo  [0]{\@secondoftwo}%
\providecommand \bibfield  [0]{\@secondoftwo}%
\providecommand \translation [1]{[#1]}%
\providecommand \BibitemOpen [0]{}%
\providecommand \bibitemStop [0]{}%
\providecommand \bibitemNoStop [0]{.\EOS\space}%
\providecommand \EOS [0]{\spacefactor3000\relax}%
\providecommand \BibitemShut  [1]{\csname bibitem#1\endcsname}%
\let\auto@bib@innerbib\@empty
\bibitem [{\citenamefont {Elvang}\ \emph {et~al.}(2019)\citenamefont {Elvang},
  \citenamefont {Hadjiantonis}, \citenamefont {Jones},\ and\ \citenamefont
  {Paranjape}}]{Elvang:2018dco}%
  \BibitemOpen
  \bibfield  {author} {\bibinfo {author} {\bibfnamefont {H.}~\bibnamefont
  {Elvang}}, \bibinfo {author} {\bibfnamefont {M.}~\bibnamefont
  {Hadjiantonis}}, \bibinfo {author} {\bibfnamefont {C.~R.}\ \bibnamefont
  {Jones}}, \ and\ \bibinfo {author} {\bibfnamefont {S.}~\bibnamefont
  {Paranjape}},\ }\href {\doibase 10.1007/JHEP01(2019)195} {\bibfield
  {journal} {\bibinfo  {journal} {JHEP}\ }\textbf {\bibinfo {volume} {01}},\
  \bibinfo {pages} {195} (\bibinfo {year} {2019})},\ \Eprint
  {http://arxiv.org/abs/1806.06079} {arXiv:1806.06079 [hep-th]} \BibitemShut
  {NoStop}%
\bibitem [{\citenamefont {Elvang}(2020)}]{Elvang:2020lue}%
  \BibitemOpen
  \bibfield  {author} {\bibinfo {author} {\bibfnamefont {H.}~\bibnamefont
  {Elvang}},\ }\href@noop {} {\  (\bibinfo {year} {2020})},\ \Eprint
  {http://arxiv.org/abs/2007.08436} {arXiv:2007.08436 [hep-th]} \BibitemShut
  {NoStop}%
\bibitem [{\citenamefont {Grzadkowski}\ \emph {et~al.}(2010)\citenamefont
  {Grzadkowski}, \citenamefont {Iskrzynski}, \citenamefont {Misiak},\ and\
  \citenamefont {Rosiek}}]{Grzadkowski:2010es}%
  \BibitemOpen
  \bibfield  {author} {\bibinfo {author} {\bibfnamefont {B.}~\bibnamefont
  {Grzadkowski}}, \bibinfo {author} {\bibfnamefont {M.}~\bibnamefont
  {Iskrzynski}}, \bibinfo {author} {\bibfnamefont {M.}~\bibnamefont {Misiak}},
  \ and\ \bibinfo {author} {\bibfnamefont {J.}~\bibnamefont {Rosiek}},\ }\href
  {\doibase 10.1007/JHEP10(2010)085} {\bibfield  {journal} {\bibinfo  {journal}
  {JHEP}\ }\textbf {\bibinfo {volume} {10}},\ \bibinfo {pages} {085} (\bibinfo
  {year} {2010})},\ \Eprint {http://arxiv.org/abs/1008.4884} {arXiv:1008.4884
  [hep-ph]} \BibitemShut {NoStop}%
\bibitem [{\citenamefont {Liao}\ and\ \citenamefont {Ma}(2017)}]{Liao:2016qyd}%
  \BibitemOpen
  \bibfield  {author} {\bibinfo {author} {\bibfnamefont {Y.}~\bibnamefont
  {Liao}}\ and\ \bibinfo {author} {\bibfnamefont {X.-D.}\ \bibnamefont {Ma}},\
  }\href {\doibase 10.1103/PhysRevD.96.015012} {\bibfield  {journal} {\bibinfo
  {journal} {Phys. Rev. D}\ }\textbf {\bibinfo {volume} {96}},\ \bibinfo
  {pages} {015012} (\bibinfo {year} {2017})},\ \Eprint
  {http://arxiv.org/abs/1612.04527} {arXiv:1612.04527 [hep-ph]} \BibitemShut
  {NoStop}%
\bibitem [{\citenamefont {Murphy}(2020)}]{Murphy:2020rsh}%
  \BibitemOpen
  \bibfield  {author} {\bibinfo {author} {\bibfnamefont {C.~W.}\ \bibnamefont
  {Murphy}},\ }\href {\doibase 10.1007/JHEP10(2020)174} {\bibfield  {journal}
  {\bibinfo  {journal} {JHEP}\ }\textbf {\bibinfo {volume} {10}},\ \bibinfo
  {pages} {174} (\bibinfo {year} {2020})},\ \Eprint
  {http://arxiv.org/abs/2005.00059} {arXiv:2005.00059 [hep-ph]} \BibitemShut
  {NoStop}%
\bibitem [{\citenamefont {Henning}\ \emph {et~al.}(2017)\citenamefont
  {Henning}, \citenamefont {Lu}, \citenamefont {Melia},\ and\ \citenamefont
  {Murayama}}]{Henning:2015alf}%
  \BibitemOpen
  \bibfield  {author} {\bibinfo {author} {\bibfnamefont {B.}~\bibnamefont
  {Henning}}, \bibinfo {author} {\bibfnamefont {X.}~\bibnamefont {Lu}},
  \bibinfo {author} {\bibfnamefont {T.}~\bibnamefont {Melia}}, \ and\ \bibinfo
  {author} {\bibfnamefont {H.}~\bibnamefont {Murayama}},\ }\href {\doibase
  10.1007/JHEP09(2019)019, 10.1007/JHEP08(2017)016} {\bibfield  {journal}
  {\bibinfo  {journal} {JHEP}\ }\textbf {\bibinfo {volume} {08}},\ \bibinfo
  {pages} {016} (\bibinfo {year} {2017})},\ \bibinfo {note} {[Erratum:
  JHEP09,019(2019)]},\ \Eprint {http://arxiv.org/abs/1512.03433}
  {arXiv:1512.03433 [hep-ph]} \BibitemShut {NoStop}%
\bibitem [{\citenamefont {Fonseca}(2019)}]{Fonseca:2019yya}%
  \BibitemOpen
  \bibfield  {author} {\bibinfo {author} {\bibfnamefont {R.~M.}\ \bibnamefont
  {Fonseca}},\ }\href@noop {} {\  (\bibinfo {year} {2019})},\ \Eprint
  {http://arxiv.org/abs/1907.12584} {arXiv:1907.12584 [hep-ph]} \BibitemShut
  {NoStop}%
\bibitem [{\citenamefont {Ma}\ \emph {et~al.}(2019)\citenamefont {Ma},
  \citenamefont {Shu},\ and\ \citenamefont {Xiao}}]{Ma:2019gtx}%
  \BibitemOpen
  \bibfield  {author} {\bibinfo {author} {\bibfnamefont {T.}~\bibnamefont
  {Ma}}, \bibinfo {author} {\bibfnamefont {J.}~\bibnamefont {Shu}}, \ and\
  \bibinfo {author} {\bibfnamefont {M.-L.}\ \bibnamefont {Xiao}},\ }\href@noop
  {} {\  (\bibinfo {year} {2019})},\ \Eprint {http://arxiv.org/abs/1902.06752}
  {arXiv:1902.06752 [hep-ph]} \BibitemShut {NoStop}%
\bibitem [{\citenamefont {Shadmi}\ and\ \citenamefont
  {Weiss}(2019)}]{Shadmi:2018xan}%
  \BibitemOpen
  \bibfield  {author} {\bibinfo {author} {\bibfnamefont {Y.}~\bibnamefont
  {Shadmi}}\ and\ \bibinfo {author} {\bibfnamefont {Y.}~\bibnamefont {Weiss}},\
  }\href {\doibase 10.1007/JHEP02(2019)165} {\bibfield  {journal} {\bibinfo
  {journal} {JHEP}\ }\textbf {\bibinfo {volume} {02}},\ \bibinfo {pages} {165}
  (\bibinfo {year} {2019})},\ \Eprint {http://arxiv.org/abs/1809.09644}
  {arXiv:1809.09644 [hep-ph]} \BibitemShut {NoStop}%
\bibitem [{\citenamefont {Li}\ \emph {et~al.}(2020{\natexlab{a}})\citenamefont
  {Li}, \citenamefont {Ren}, \citenamefont {Shu}, \citenamefont {Xiao},
  \citenamefont {Yu},\ and\ \citenamefont {Zheng}}]{Li:2020gnx}%
  \BibitemOpen
  \bibfield  {author} {\bibinfo {author} {\bibfnamefont {H.-L.}\ \bibnamefont
  {Li}}, \bibinfo {author} {\bibfnamefont {Z.}~\bibnamefont {Ren}}, \bibinfo
  {author} {\bibfnamefont {J.}~\bibnamefont {Shu}}, \bibinfo {author}
  {\bibfnamefont {M.-L.}\ \bibnamefont {Xiao}}, \bibinfo {author}
  {\bibfnamefont {J.-H.}\ \bibnamefont {Yu}}, \ and\ \bibinfo {author}
  {\bibfnamefont {Y.-H.}\ \bibnamefont {Zheng}},\ }\href@noop {} {\  (\bibinfo
  {year} {2020}{\natexlab{a}})},\ \Eprint {http://arxiv.org/abs/2005.00008}
  {arXiv:2005.00008 [hep-ph]} \BibitemShut {NoStop}%
\bibitem [{\citenamefont {Li}\ \emph {et~al.}(2020{\natexlab{b}})\citenamefont
  {Li}, \citenamefont {Ren}, \citenamefont {Xiao}, \citenamefont {Yu},\ and\
  \citenamefont {Zheng}}]{Li:2020xlh}%
  \BibitemOpen
  \bibfield  {author} {\bibinfo {author} {\bibfnamefont {H.-L.}\ \bibnamefont
  {Li}}, \bibinfo {author} {\bibfnamefont {Z.}~\bibnamefont {Ren}}, \bibinfo
  {author} {\bibfnamefont {M.-L.}\ \bibnamefont {Xiao}}, \bibinfo {author}
  {\bibfnamefont {J.-H.}\ \bibnamefont {Yu}}, \ and\ \bibinfo {author}
  {\bibfnamefont {Y.-H.}\ \bibnamefont {Zheng}},\ }\href@noop {} {\  (\bibinfo
  {year} {2020}{\natexlab{b}})},\ \Eprint {http://arxiv.org/abs/2007.07899}
  {arXiv:2007.07899 [hep-ph]} \BibitemShut {NoStop}%
\bibitem [{\citenamefont {Henning}\ and\ \citenamefont
  {Melia}(2019{\natexlab{a}})}]{Henning:2019enq}%
  \BibitemOpen
  \bibfield  {author} {\bibinfo {author} {\bibfnamefont {B.}~\bibnamefont
  {Henning}}\ and\ \bibinfo {author} {\bibfnamefont {T.}~\bibnamefont
  {Melia}},\ }\href {\doibase 10.1103/PhysRevD.100.016015} {\bibfield
  {journal} {\bibinfo  {journal} {Phys. Rev.}\ }\textbf {\bibinfo {volume}
  {D100}},\ \bibinfo {pages} {016015} (\bibinfo {year} {2019}{\natexlab{a}})},\
  \Eprint {http://arxiv.org/abs/1902.06754} {arXiv:1902.06754 [hep-ph]}
  \BibitemShut {NoStop}%
\bibitem [{\citenamefont {Henning}\ and\ \citenamefont
  {Melia}(2019{\natexlab{b}})}]{Henning:2019mcv}%
  \BibitemOpen
  \bibfield  {author} {\bibinfo {author} {\bibfnamefont {B.}~\bibnamefont
  {Henning}}\ and\ \bibinfo {author} {\bibfnamefont {T.}~\bibnamefont
  {Melia}},\ }\href@noop {} {\  (\bibinfo {year} {2019}{\natexlab{b}})},\
  \Eprint {http://arxiv.org/abs/1902.06747} {arXiv:1902.06747 [hep-th]}
  \BibitemShut {NoStop}%
\bibitem [{\citenamefont {Li}\ \emph {et~al.}(2020{\natexlab{c}})\citenamefont
  {Li}, \citenamefont {Ren}, \citenamefont {Xiao}, \citenamefont {Yu},\ and\
  \citenamefont {Zheng}}]{Li:2020tsi}%
  \BibitemOpen
  \bibfield  {author} {\bibinfo {author} {\bibfnamefont {H.-L.}\ \bibnamefont
  {Li}}, \bibinfo {author} {\bibfnamefont {Z.}~\bibnamefont {Ren}}, \bibinfo
  {author} {\bibfnamefont {M.-L.}\ \bibnamefont {Xiao}}, \bibinfo {author}
  {\bibfnamefont {J.-H.}\ \bibnamefont {Yu}}, \ and\ \bibinfo {author}
  {\bibfnamefont {Y.-H.}\ \bibnamefont {Zheng}},\ }\href@noop {} {\  (\bibinfo
  {year} {2020}{\natexlab{c}})},\ \Eprint {http://arxiv.org/abs/2012.09188}
  {arXiv:2012.09188 [hep-ph]} \BibitemShut {NoStop}%
\bibitem [{\citenamefont {Li}\ \emph {et~al.}()\citenamefont {Li},
  \citenamefont {Ren}, \citenamefont {Xiao}, \citenamefont {Yu},\ and\
  \citenamefont {Zheng}}]{nuSMEFT}%
  \BibitemOpen
  \bibfield  {author} {\bibinfo {author} {\bibfnamefont {H.-L.}\ \bibnamefont
  {Li}}, \bibinfo {author} {\bibfnamefont {Z.}~\bibnamefont {Ren}}, \bibinfo
  {author} {\bibfnamefont {M.-L.}\ \bibnamefont {Xiao}}, \bibinfo {author}
  {\bibfnamefont {J.-H.}\ \bibnamefont {Yu}}, \ and\ \bibinfo {author}
  {\bibfnamefont {Y.-H.}\ \bibnamefont {Zheng}},\ }\href@noop {} {\ }\Eprint
  {http://arxiv.org/abs/In Preparation} {arXiv:In Preparation [hep-ph]}
  \BibitemShut {NoStop}%
\bibitem [{\citenamefont {Jiang}\ \emph {et~al.}(2020)\citenamefont {Jiang},
  \citenamefont {Shu}, \citenamefont {Xiao},\ and\ \citenamefont
  {Zheng}}]{Jiang:2020sdh}%
  \BibitemOpen
  \bibfield  {author} {\bibinfo {author} {\bibfnamefont {M.}~\bibnamefont
  {Jiang}}, \bibinfo {author} {\bibfnamefont {J.}~\bibnamefont {Shu}}, \bibinfo
  {author} {\bibfnamefont {M.-L.}\ \bibnamefont {Xiao}}, \ and\ \bibinfo
  {author} {\bibfnamefont {Y.-H.}\ \bibnamefont {Zheng}},\ }\href@noop {} {\
  (\bibinfo {year} {2020})},\ \Eprint {http://arxiv.org/abs/2001.04481}
  {arXiv:2001.04481 [hep-ph]} \BibitemShut {NoStop}%
\bibitem [{Note1()}]{Note1}%
  \BibitemOpen
  \bibinfo {note} {Regarding matrix element between physical states, the EOM is
  equivalent to zero.}\BibitemShut {Stop}%
\bibitem [{\citenamefont {Liao}\ and\ \citenamefont {Ma}(2019)}]{Liao:2019tep}%
  \BibitemOpen
  \bibfield  {author} {\bibinfo {author} {\bibfnamefont {Y.}~\bibnamefont
  {Liao}}\ and\ \bibinfo {author} {\bibfnamefont {X.-D.}\ \bibnamefont {Ma}},\
  }\href {\doibase 10.1007/JHEP03(2019)179} {\bibfield  {journal} {\bibinfo
  {journal} {JHEP}\ }\textbf {\bibinfo {volume} {03}},\ \bibinfo {pages} {179}
  (\bibinfo {year} {2019})},\ \Eprint {http://arxiv.org/abs/1901.10302}
  {arXiv:1901.10302 [hep-ph]} \BibitemShut {NoStop}%
\bibitem [{Note2()}]{Note2}%
  \BibitemOpen
  \bibinfo {note} {The usual Littlewood-Richardson rule is for Young diagrams
  compared to ours for Young tableaux.}\BibitemShut {Stop}%
\bibitem [{\citenamefont {Ma}(2008)}]{book:Ma}%
  \BibitemOpen
  \bibfield  {author} {\bibinfo {author} {\bibfnamefont {Z.-Q.}\ \bibnamefont
  {Ma}},\ }\href
  {http://gen.lib.rus.ec/book/index.php?md5=35F872F9CC024EC1D0F7B418F4062F2E}
  {\emph {\bibinfo {title} {Group Theory for Physicists}}}\ (\bibinfo
  {publisher} {WS},\ \bibinfo {year} {2008})\BibitemShut {NoStop}%
\bibitem [{Note3()}]{Note3}%
  \BibitemOpen
  \bibinfo {note} {The \protect \emph {term} defined in our framework differ
  from the same terminology used in other literature such as \cite
  {Fonseca:2019yya,Murphy:2020rsh}. In those context, a term stands for a
  flavor tensor of operators with redundancies called flavor relations, which
  may consist of several irreducible representations, i.e. the direct sum of
  several terms in our sense. The number of terms is ambiguous as given in
  \cite {Fonseca:2019yya}, and even the upper bound is less than that in our
  interpretation.}\BibitemShut {Stop}%
\bibitem [{Note4()}]{Note4}%
  \BibitemOpen
  \bibinfo {note} {When taking the inverse of $\protect \mathcal {K}^{\protect
  \rm pj}$, we are using the full $\protect \mathbbm {d}\times \protect
  \mathbbm {d}$ matrix, even if some of them are not relevant in the complete
  basis of flavor-specified operators.}\BibitemShut {Stop}%
\bibitem [{\citenamefont {Nussinov}\ and\ \citenamefont
  {Shrock}(2002)}]{Nussinov:2001rb}%
  \BibitemOpen
  \bibfield  {author} {\bibinfo {author} {\bibfnamefont {S.}~\bibnamefont
  {Nussinov}}\ and\ \bibinfo {author} {\bibfnamefont {R.}~\bibnamefont
  {Shrock}},\ }\href {\doibase 10.1103/PhysRevLett.88.171601} {\bibfield
  {journal} {\bibinfo  {journal} {Phys. Rev. Lett.}\ }\textbf {\bibinfo
  {volume} {88}},\ \bibinfo {pages} {171601} (\bibinfo {year} {2002})},\
  \Eprint {http://arxiv.org/abs/hep-ph/0112337} {arXiv:hep-ph/0112337}
  \BibitemShut {NoStop}%
\bibitem [{\citenamefont {Babu}\ \emph {et~al.}(2006)\citenamefont {Babu},
  \citenamefont {Mohapatra},\ and\ \citenamefont {Nasri}}]{Babu:2006xc}%
  \BibitemOpen
  \bibfield  {author} {\bibinfo {author} {\bibfnamefont {K.}~\bibnamefont
  {Babu}}, \bibinfo {author} {\bibfnamefont {R.}~\bibnamefont {Mohapatra}}, \
  and\ \bibinfo {author} {\bibfnamefont {S.}~\bibnamefont {Nasri}},\ }\href
  {\doibase 10.1103/PhysRevLett.97.131301} {\bibfield  {journal} {\bibinfo
  {journal} {Phys. Rev. Lett.}\ }\textbf {\bibinfo {volume} {97}},\ \bibinfo
  {pages} {131301} (\bibinfo {year} {2006})},\ \Eprint
  {http://arxiv.org/abs/hep-ph/0606144} {arXiv:hep-ph/0606144} \BibitemShut
  {NoStop}%
\bibitem [{\citenamefont {Assad}\ \emph {et~al.}(2018)\citenamefont {Assad},
  \citenamefont {Fornal},\ and\ \citenamefont {Grinstein}}]{Assad:2017iib}%
  \BibitemOpen
  \bibfield  {author} {\bibinfo {author} {\bibfnamefont {N.}~\bibnamefont
  {Assad}}, \bibinfo {author} {\bibfnamefont {B.}~\bibnamefont {Fornal}}, \
  and\ \bibinfo {author} {\bibfnamefont {B.}~\bibnamefont {Grinstein}},\ }\href
  {\doibase 10.1016/j.physletb.2017.12.042} {\bibfield  {journal} {\bibinfo
  {journal} {Phys. Lett. B}\ }\textbf {\bibinfo {volume} {777}},\ \bibinfo
  {pages} {324} (\bibinfo {year} {2018})},\ \Eprint
  {http://arxiv.org/abs/1708.06350} {arXiv:1708.06350 [hep-ph]} \BibitemShut
  {NoStop}%
\bibitem [{\citenamefont {Baratella}\ \emph {et~al.}(2020)\citenamefont
  {Baratella}, \citenamefont {Fernandez}, \citenamefont {von Harling},\ and\
  \citenamefont {Pomarol}}]{Baratella:2020dvw}%
  \BibitemOpen
  \bibfield  {author} {\bibinfo {author} {\bibfnamefont {P.}~\bibnamefont
  {Baratella}}, \bibinfo {author} {\bibfnamefont {C.}~\bibnamefont
  {Fernandez}}, \bibinfo {author} {\bibfnamefont {B.}~\bibnamefont {von
  Harling}}, \ and\ \bibinfo {author} {\bibfnamefont {A.}~\bibnamefont
  {Pomarol}},\ }\href@noop {} {\  (\bibinfo {year} {2020})},\ \Eprint
  {http://arxiv.org/abs/2010.13809} {arXiv:2010.13809 [hep-ph]} \BibitemShut
  {NoStop}%
\end{thebibliography}%

\end{document}